\documentstyle{elsart}
\input{epsf}
\begin{document}

\begin{frontmatter}
\title{The Radio-Jet X-Ray Binaries} 
\author[Brighton]{R.P. Fender,}
\author[OU]{S.J. Bell Burnell} 
\author[NRL]{ \& E.B. Waltman}

\address[Brighton]{Astronomy Centre, University of Sussex, Falmer,
Brighton, BN1 9QH U.K.}
\address[OU]{Department of Physics, The Open University, Walton Hall,
Milton Keynes MK7 6AA U.K.}
\address[NRL]{Remote Sensing Division, Naval Research Laboratory, Code
7210, Washington, DC 20375-5351 U.S.A.}

\begin{abstract}
We review the observational properties, from radio to X-rays, 
of the eight systems for
which there is strong evidence for the formation of a radio jet
in a X-ray binary system. 
\end{abstract}
\end{frontmatter}

\section{Introduction}

Radio jets from an X-ray binary were first discovered from SS433 by Spencer
\cite{Spencer1979} and subsequently mapped by Hjellming \& Johnston
\cite{Hjellming1981a} with the VLA.  The jets had
earlier been predicted on the basis of Doppler-shifting emission lines
(e.g. \cite{Margon1980}).

Over the past fifteen years the class of radio-jet X-ray binaries
(RJXRB) has expanded slowly to include approximately eight sources for
which there is direct observational evidence of jet formation.  These
systems, in approximate order of jet discovery, are :

\begin{itemize}

\item{Cyg X-3, for which an expansion velocity of $\sim 0.3$c has been
repeatedly measured during outbursts (e.g. \cite{Schalinski1995}) but
in which no observations have to date been able to resolve the motion
of individual plasmons.}

\item{1E1740.7-2942 \& GRS 1758-258, which while not definitely
established as X-ray binaries, are bright hard X-ray sources near the
galactic centre with spectacular arcmin-scale radio jets
\cite{Mirabel1994b}.}

\item{LSI+61$^{\circ}$ 303, a periodic radio flaring source with
measured expansion following outburst \cite{Massi1993}.}

\item{Cir X-1, another periodic flaring source which is embedded in a
synchrotron structure with associated radio jets \cite{Stewart1993}.}

\item{GRS 1915+105, a spectacular X-ray and radio transient which
mapping with VLA has revealed to possess relativistic jets with
apparent superluminal motion, implying true velocities of $\sim 0.9$c
\cite{Mirabel1994a}.}

\item{GRO J1655-40, a second superluminal X-ray transient, with $\sim
0.9$c jets resolved by southern hemisphere VLBI \cite{Tingay1995} and
the VLA \cite{Hjellming1995b}.}

\end{itemize}

Only SS 433 and the two superluminal transients indisputably possess
collimated outflows, while 1E 1740.7-2942 \& GRS 1758-258 are not
actually established as binary systems (the binarity of GRS 1915+105
has not been definitively established but seems highly likely) -- in
our opinion the order of reality of the RJXRB classification (with the
first three beyond doubt) is : SS 433, GRS 1915+105, GRO J1655-40, Cyg
X-3, Cir X-1, LSI+61$^{\circ}$ 303, 1E 1740.7-2942 \& GRS 1758-258. We
do not include in the list the sources GT 2318+620 \cite{Taylor1991}
and GX 1+4 \cite{Manchanda1993} for which RJXRB status has been
claimed but the evidence is at present uncertain.

Table 1 summarises the overall properties of
the RJXRBs. In the following sections we review the observational
properties of these sources, looking for any common properties which
might link them. 

\section{A multiwavelength comparison}

\begin{table*}
\caption[]{Radio-jet X-ray sources : system properties}
\begin{flushleft}
\begin{tabular}{lllll}
\noalign{\smallskip}
\hline
\noalign{\smallskip}
Source & Spectral class & Compact & Periodicities & Distance \\ 
       & of companion   & object  &(`o' = orbital)& (kpc)     \\
\hline
SS433 & OB (?) & ? & 13 d (o), 164 d & 8 \\
GRS 1915+105 & Red giant/dwarf ? & black hole & & 10 -- 12.5 \\
GRO J1655-40  & F or G & black hole & 2.6 d (o) & 3 -- 6 \\
Cyg X-3 & W-R (?) & ? & 4.8 h (o)& 8.5--12 \\
Cir X-1 & MS & neutron star & 16.6 d (o)& $\geq 6.5$\\
LSI+61$^{\circ}$ 303 & Be & ? & 26.5 d (o), 4 yr (?) & $\sim 2$\\
1E 1740.7-2942 & binary ? & black hole ? & & 8.5 (g.c.)\\
GRS 1758-58 & binary ? & ? & & 8.5 (g.c.)\\
\noalign{\smallskip}
\hline
\end{tabular}
\end{flushleft}
\end{table*}

\subsection{Radio}

\begin{table*}
\begin{center}
\caption{Summary of Radio and (sub)mm observations of radio-jet XRBs.}
\small
\begin{tabular}{llll}
\hline 
& F$_{cm}$ quiescent/ & F$_{mm}$ quiescent/& notes \\ 
& brightest flare (Jy) & brightest flare (Jy) & \\ 
\hline 
SS433 & $\sim 0.5$ / $\geq 10$ & $\sim 0.1$ & (1) \\
GRS 1915+105 & $\sim 0.01$ / $\geq 1$ & $< 0.01$& (2) \\
GRO J1655-40 & $< 0.01$ / $\geq 7$ & -- & (2) \\
Cyg X-3 & $\sim 0.05$ / $\geq 20$ & $\sim 0.05$ / $\geq 3$ & (3) \\
Cir X-1 & $< 0.1$ / $>3.0$ & -- & (4) \\
LSI +61$^{\circ}$ 303 & $\sim 0.03$ / $\geq 0.5$ & $\sim 0.01$ & (5)\\ 
1E 1740.7-2942 & ($< 1$ / 5) $\times 10{-3}$ & $< 0.11$ & \\
GRS 1758-258 & ($< 1$ / 5) $\times 10{-3}$ & $< 0.09$ & \\ 
\hline
\end{tabular}
\end{center}
\vspace*{0.5cm}
\footnotesize{(1) cm -- mm optically thin (2) Radio flaring correlated
with X-ray activity (3) Radio flaring correlated with X-ray activity,
mm excess (4) Modulated at 16.6 d orbital period, decline since 1970s
(5) Modulated at 26.5 d orbital period, cm -- mm optically thin } 
\normalsize
\end{table*}

\begin{figure*}
\centering
\setlength{\unitlength}{1cm}
\begin{picture}(10,10)(0,0)
\put(0,0){\includegraphics{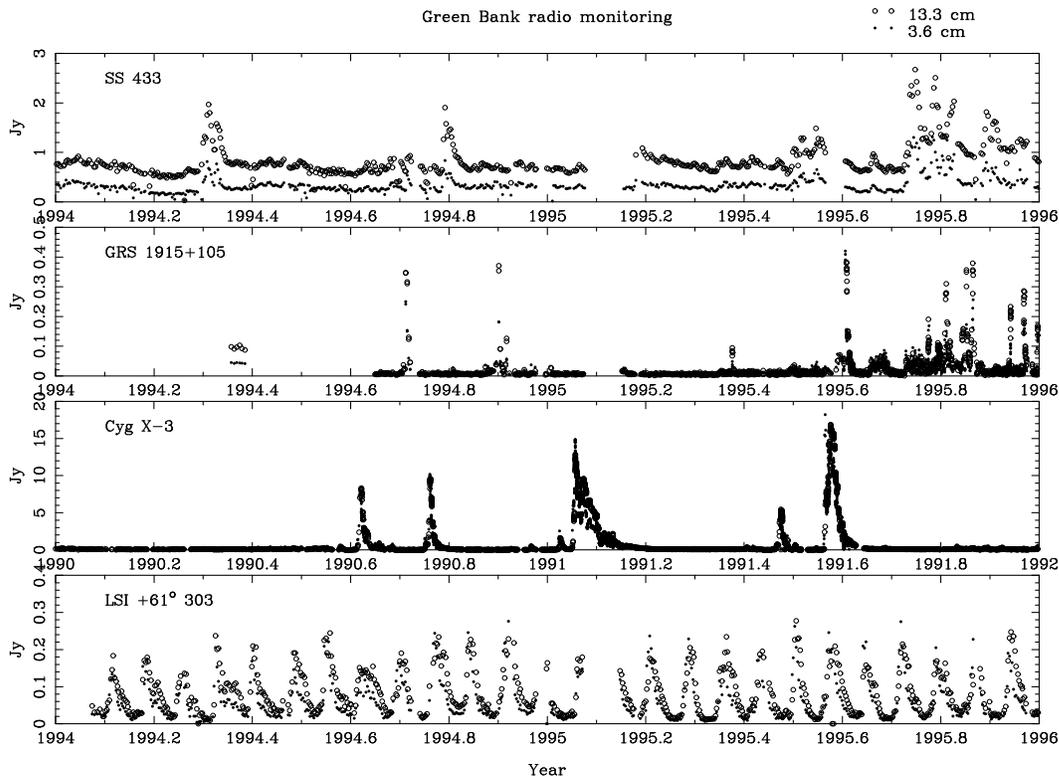}}
\end{picture}
%\vspace{10cm}
%\leavevmode\epsffile{four.ps}
\caption{Two-year radio flux histories of SS 433, GRS 1915+105, Cyg
X-3 \& LSI+61$^{\circ}$ 303 from the NRL-Green Bank monitoring
program. Note that the period for Cyg X-3 is different from that of
the others in order to highlight some flaring activity.}
\end{figure*}

All the sources are radio-bright, unusual in itself for X-ray binaries
(some $\sim 25$ of the $\sim 200$ known X-ray binaries have been found
to have detectable radio emission -- see
e.g. \cite{Hjellming1995a}). In all cases the radio emission mechanism
is dominated by non-thermal synchrotron emission, characterised by
very high ($> 10^8$ K) brightness temperatures and a negative spectral
index $\alpha = \Delta \log S / \Delta \log \nu$.

Fig 1 plots two-year flux histories of SS 433, GRS 1915+105, Cyg X-3
and LSI+61$^{\circ}$ 303 as observed during the (now terminated due to
lack of funding) NRL-Green Bank monitoring program; table 2 summarises
the radio properties of the RJXRBs.

SS 433 is a persistent bright radio source undergoing occasional
outbursts during which its flux may increase by up to a factor of ten
or so (e.g. \cite{Bursov1995}). The resolved radio plasmons trace out
a corkscrew pattern on the sky reflecting (probably) the precession of
the accretion disc \cite{Hjellming1981b}.  Large radio outbursts may
correspond to the ejection of particularly bright plasmons from the
system.  The emission from the source is consistently optically thin
(clear from the near-constant separation of the 13.3 cm \& 3.6 cm
light curves for SS 433 in Fig 1), and the optically thin tail has
been measured out to mm wavelengths \cite{Tsutsumi1996}.

The radio emissions from GRS 1915+105 and GRO J1655-40 share many
common properties. Both systems are below the detection limits of
current telescopes for the majority of the time between outbursts, and
during the outbursts the flux density can increase by a factor
of $> 1000$. It is during these outburst periods, correlated (to a certain
extent) with X-ray flaring (e.g. \cite{Foster1996,Hjellming1995b})
that individual plasmons are tracked moving away from the systems at
velocities of order 0.9 c \cite{Mirabel1994a,Tingay1995}. The spectral
index of emission from both sources indicates varying degrees of
opacity, with absorption particularly prominent in the early stages of
outbursts.  Upper limits to the (sub)millimetre emission from GRS
1915+105 \cite{Tsutsumi1996} are consistent with an optically thin
spectrum.  High time resolution observations of GRS 1915+105 have
revealed radio QPO with periods in the range 20 -- 120 min
\cite{Pooley1996a,Pooley1996b}.

Cyg X-3 is a persistent bright radio source, occasionally undergoing
huge outbursts when its radio flux density can increase by factors $>
50$ on timescales of days (e.g. \cite{Gregory1972} et seq). At these
times the source is observed to become an expanding radio source with
expansion velocity $\sim 0.3$c (e.g. \cite{Schalinski1995}).  Prior to
major outbursts, the radio flux from the source is often observed to
drop to very low levels (e.g. \cite{Waltman1994}).  Quiescent emission
in Cyg X-3 from cm -- mm is typically flat with a significant excess
at mm wavelengths \cite{Fender1995} indicative of absorption. During
flaring sequences the emission typically progresses from absorbed to
optically thin both within individual flare events and along the
flaring sequence (e.g. \cite{Fender1997a}).

About the radio emission from Cir X-1, less is known. The source was
observed several times in the 1970s to undergo radio flaring once
every 16.6 days but has since then been declining steadily in radio
luminosity (G. Nicholson, private communication). Mapping of the
system has revealed it to be embedded within a synchrotron nebula, with
swept-back jets suggesting ejection from the nearby SNR G321.9-0.3
\cite{Stewart1993}.

LSI+61$^{\circ}$ 303, like Cir X-1, is a periodically flaring source,
this time once every 26.5 days (presumably orbital in origin)
(e.g. \cite{Ray1996}).  Radio mapping following one outburst suggests
an expansion velocity of several 1000 km s$^{-1}$ \cite{Massi1993}.
During radio flares the source can increase its flux by a factor of 5
-- 10, similar to SS 433.

Both 1E 1740.7-2942 \& GRS 1758-258 have associated arcmin-scale radio
jets \cite{Mirabel1994b}, and while radio monitoring has been sporadic
due to their relative faintness, they are clearly variable sources
\cite{Marti1993}.

\subsection{Infrared}

Table 3 summarises the infrared properties of the RJXRBs.

\begin{table*}
\caption{Summary of infrared observations of radio-jet XRBs}
\begin{center}
\small
\begin{tabular}{llll}
\hline
& F$_{2.2 \mu m}$ quiescent/ & spectral features & notes \\
& brightest flare (mJy)  & & \\
\hline
SS433 & $> 50$ & H, He I & (1)\\
& & (Doppler shifted)& \\
GRS 1915+105 & $\leq 1$ / $\geq 4$ & H, He I (flaring) & (2)\\
GRO J1655-40  & -- & -- & \\
Cyg X-3 & $\sim 12$ / $\geq 50$ & He I, He II, N & (3) \\
& & (Wolf-Rayet-like) & \\
Cir X-1 & $\sim 15$ & -- & (4)\\
LSI +61 $^{\circ}$ 303 & $\sim 400$ & -- & (5)\\
1E 1740.7-2942 & $< 0.1$ & -- & (6)\\
GRS 1758-258 & $< 0.1$ & -- & (6)\\
\hline
\end{tabular}
\end{center}
\vspace*{0.5cm} 
\footnotesize{(1) Modulated at 13 d orbital and 164 d
disc precession periods (2) $> 1$ mag variability at JHK (3) modulated
at 4.8 h orbital period, rapid flares, no H in spectrum (4) large
broad flare once per 16.6 d orbit (5) modulated at 26.5 d orbital
period (6) K$>$ 17} \\
\normalsize
\end{table*}

SS 433 has been observed in the infrared both photometrically and
spectroscopically \cite{Kodaira1985,Thompson1979}.  Spectral studies
reveal stationary and Doppler-shifted lines (see section 2.3) while
photometry reveals a modulation at the 13 d orbital period which is
itself modulated in shape with the 164 d precession period. This
effect is interpreted by Kodaira et al \cite{Kodaira1985} as arising
in infrared emission from an accretion disc of which different aspects
are on view at different disc precession phases. Infrared imaging with
the IRAS satellite has revealed a series of `knots' which may be
associated with the jets \cite{Wang1990}.

GRS 1915+105 has a variable infrared counterpart in the JHK bands
(e.g. \cite{Chaty1996}), but no periodicities have been found in the
modulation. Infrared spectroscopy (e.g. \cite{CastroTirada1996}) has
revealed H \& He emission lines which appear to be more prominent when
the system is in an active state.  High-resolution infrared imaging
may have resolved the jet in GRS 1915+105 \cite{Sams1996}.

Infrared observations of GRO J1655-40 have been few due to a
relatively bright optical counterpart, but the infrared flux has been
seen to brighten during radio outburst and there may be evidence for
ellipsoidal modulation of the secondary \cite{Callanan1997}.

Cyg X-3 is a bright infrared source, which modulates in the IJHKL
bands at the 4.8 hr (presumed) orbital period (e.g. \cite{Mason1986}).
Superimposed upon this modulation are often observed rapid flare
events which may arise in hot ($10^6$ K) gas associated with the
disc/jet \cite{Fender1996}.  Infrared spectroscopy of Cyg X-3 has
revealed broad Doppler-shifted emission lines reminiscent of a
Wolf-Rayet star, suggesting that Cyg X-3 may be the first W-R +
compact object system identified \cite{vKerkwijk1992}.

Cir X-1 has been observed to flare in the infrared at the same period
as the radio emission \cite{Glass1994}, but few recent observations
have been made.

There are claims that LSI+61$^{\circ}$ 303 modulates in the infrared
at the 26.5 d period \cite{Paredes1994}, but these remain
controversial.

No infrared counterparts have been identified for 1E 1740.7-2942 or
GRS 1758-258, to a limiting magnitude of $\sim 17$
\cite{Mirabel1994b}.

\subsection{Optical}

Table 4 summarises the optical properties of the RJXRBs.

\begin{table*}
\caption{Summary of optical properties of radio-jet XRBs}
\begin{center}
\small
\begin{tabular}{llll}
\hline
& V mag quiescent/ & spectral features& notes \\
& flaring & & \\
\hline
SS433 & 14.2 & H, He, C, N, Fe & (1) \\
GRS 1915+105 & -- & -- & (2)\\
GRO J1655-40 & $\geq 17$ / $\leq 14$ & H, He I, He II, N III& \\
Cyg X-3 & $> 24$ & --  & (3)\\
Cir X-1 & 20.6 & H$\alpha$ & (4)\\
LSI +61 $^{\circ}$ 303 & 10.7 & H, He, Si & (5)\\
1E 1740.7-2942 & -- & -- & (6)\\
GRS 1758-258 & -- & -- & (6) \\
\hline
\end{tabular}
\end{center}
\vspace*{0.5cm}
\footnotesize{(1) Stationary and Doppler-shifted lines, few
photospheric features (2) I$\sim 24$, R $> 21$ (3) I$\sim 21$, R$\sim
24$ (4) possible previous source confusion (5) `Shell' H$\alpha$ \&
H$\beta$ (6) I $> 21$} 
\normalsize
\end{table*}

SS 433 was originally identified as a strong H$\alpha$ emission-line
star (hence its inclusion in the SS catalogue). Optical spectra of the
source reveal many emission lines, all of which can be classified as
either `stationary' or `moving', the latter with varying
Doppler-shifts and associated with the jets (e.g. \cite{Margon1984}).
Photospheric features of the companion star are very difficult to
disentangle, but may be reminiscent of an OB or W-R type star (though
note there is still plenty of hydrogen in the SS 433 system). No
periodically Doppler-shifting emission lines have ever been seen in
another X-ray binary, and so SS 433 remains the only source for which
there is direct evidence for the ejection of {\em protonic} matter at
near-relativistic velocities.

GRO J1655-40 has a relatively bright, variable optical counterpart
\cite{Bailyn1995}. Spectroscopy of the source has revealed strong
emission lines, particularly during outburst, and absorption line
radial velocity studies have revealed a 2.6 d orbit and a mass
function convincingly implying a black hole as the compact object
\cite{Bailyn1995}.  Photometric variations and anomalies in the radial
velocity curve around conjunction may also suggest an eclipse.

Cir X-1 is difficult to observe in the optical not only because it is
a faint source but also because it lies within 2 arcsec of two other
stars \cite{Moneti1992}. Apart from H$\alpha$ emission
(e.g. \cite{Nicholson1980}) little is known about the optical
counterpart to the source.

LSI+61$^{\circ}$ 303 has a bright optical Be-type companion star which
shows photospheric absorption features as well as H$\alpha$ and
H$\beta$ emission from the circumstellar disc (I. A. Steele, private
communication). Attempts to determine a radial velocity curve at the
26.5 d radio period have had little success however and a good mass
function has not been determined. There may be a photometric
modulation at the radio period \cite{Paredes1994}.

GRS 1915+105, Cyg X-3, 1E 1740.7-2942 \& GRS 1758-258 have no optical
counterparts.

\subsection{X-ray}

Table 5 summarises the X-ray properties of the RJXRBs; Fig 2 shows
100-day soft X-ray flux monitoring of all of the sources with the XTE
ASM.

\begin{figure*}
\begin{center}
\setlength{\unitlength}{1cm}
\begin{picture}(10,20)(0,0)
\put(0,0){\includegraphics{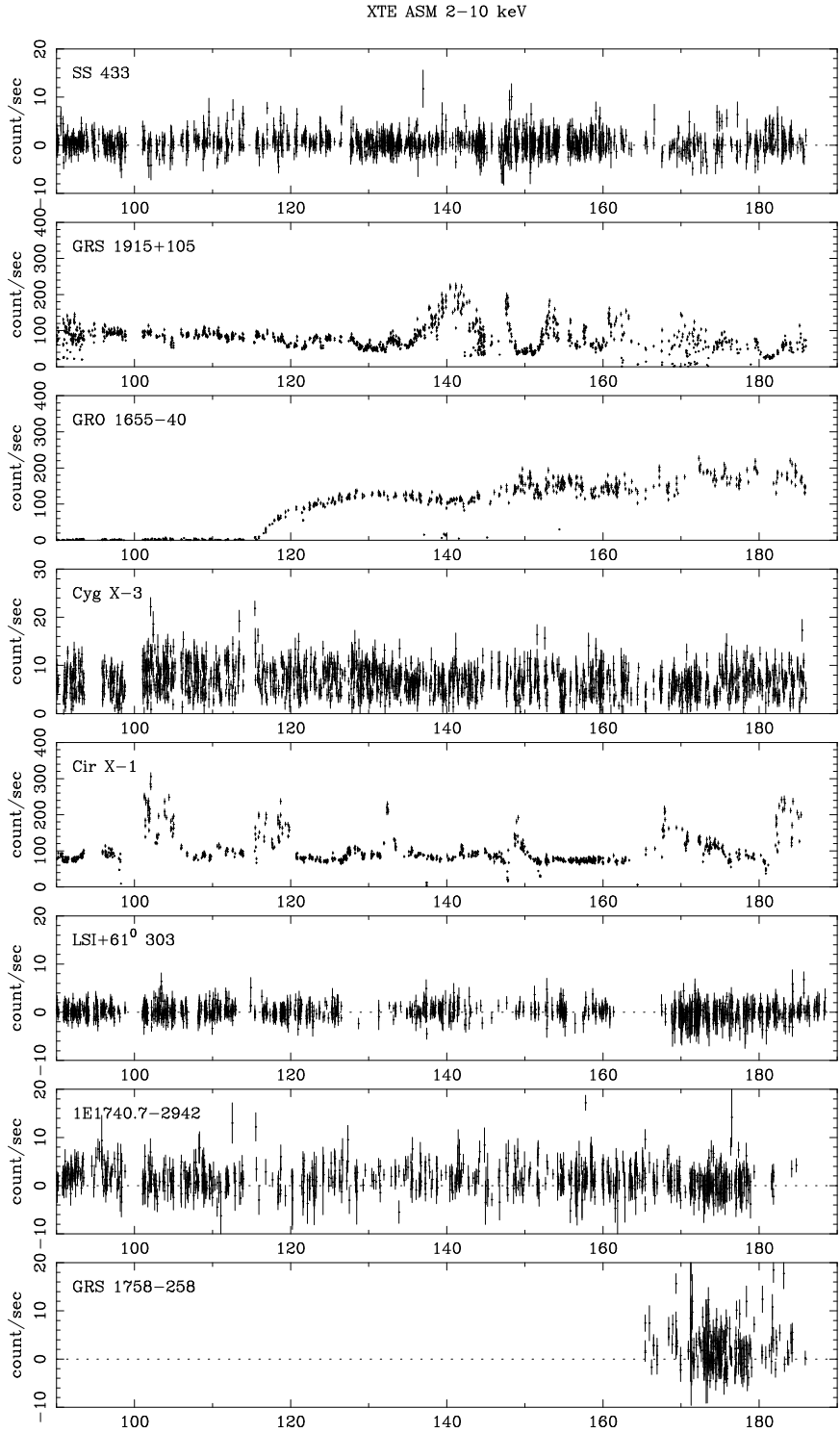}}
\end{picture}
%\vspace{20cm}
%leavevmode\epsffile{xte.ps}
\caption{100 days of flux monitoring of the RJXB in soft X-rays with
the XTE ASM (quick-look results provided by the ASM/RXTE team). Note
the rapid fluctuations in the flux from GRS 1915+105, the `flaring'
behaviour of Cir X-1, and the rapid `switch on' in the flux from GRO
J1655-40.}
\end{center}
\end{figure*}

\begin{table*}
\caption{Summary of X-ray properties of radio-jet XRBs}
\begin{center}
\small
\begin{tabular}{llll}
\hline
& X-ray luminosity in & spectral features & comments \\
& low/high states (erg s$^{-1}$) & & \\
\hline
SS433 & $10^{35} / 10^{36}$ & Fe & (1)\\
GRS 1915+105 & $ <10^{37} / > 10^{38}$ & Fe absorption & (2)\\ 
GRO J1655-40 & $<10^{37} /> 10^{38}$ & Fe absorption & (2)\\
Cyg X-3 & $10^{37} / 10^{38}$ & Fe, Si, Ne & (3) \\
Cir X-1 & $10^{37} / 10^{38}$ & & (4)\\
LSI +61 $^{\circ}$ 303 & $10^{33} / 10^{34}$ & & (5)\\
1E 1740.7-2942 & $\geq 10^{36}$ & 511 keV e$^+$e$^-$ line ? & (6)\\ 
GRS 1758-258 & $\geq 10^{36}$ & & (6)\\
\hline
\end{tabular}
\end{center}
\vspace*{0.5cm}
\footnotesize{(1) X-ray jets (2) X-ray outbursts correlated with radio,
hard X-ray tail (3) hardness anticorrelated with brightness,
modulation at 4.8 h orbital period (4) type I bursts, QPOs, `Atoll'
source (5) X-ray flares once per 26.6 d orbit (6) bright in hard
X-rays}
\normalsize
\end{table*}

SS 433 is a weak X-ray source by the standards of X-ray binaries, with
a luminosity of $\sim 10^{35}$ erg s$^{-1}$. However, while the
brightness and flux variations of the source may not be very dramatic,
it does possess spectacular X-ray jets from which up to 10\% of the
soft X-ray emission of the source may arise, which appear to confirm
the connection with the surrounding SNR W50 and place a lower limit on
the lifetime of the jet phenomenon of several 1000 yr \cite{Watson1983}.

GRS 1915+105 \& GRO J1655-40 share many similar X-ray properties as
bright transient sources. Both tend to rise from obscurity to amongst
the most luminous X-ray sources in the galaxy on short timescales (see
Fig 2 for a nice example of GRO J1655-40 `switching on').  ASCA X-ray
spectra of both sources also reveal iron absorption features in the
range 6 -- 8 keV \cite{Ebisawa1996}.  Greiner, Predehl \& Pohl
\cite{Greiner1995} have discovered a dust scattering halo around GRO
J1655-40 with ROSAT. High time resolution observations of GRS 1915+105
with XTE reveal very rapid and dramatic flux changes at about the same
time as radio QPO were being observed \cite{Greiner1997,Fender1997b}.

Cyg X-3 is a bright and persistent X-ray source, varying generally by
no more than a factor of ten in X-ray luminosity
(e.g. \cite{Priedhorsky1986}), which is anticorrelated with X-ray
spectral hardness (e.g. \cite{Hermsen1987}).  Both soft and hard
X-rays are modulated at the 4.8 h period, {\em in phase} with the
infrared modulation \cite{Mason1986,Matz1996}.  X-ray spectra of the
source reveal a host of emission features, some of which modulate in
strength and width in {\em antiphase} with the continuum
\cite{Kitamoto1994}, and which require a nebular origin for much of
the X-ray emission \cite{Liedahl1996}. ROSAT observations have
revealed a dust scattering halo around Cyg X-3 \cite{Predehl1995}, and
a larger X-ray scattering structure may also exist \cite{Kafuku1994}.

Cir X-1 is another persistent bright source, displaying type I bursts,
QPO and movement in the X-ray colour-colour diagram characteristic of
an `atoll' low-mass XRB (e.g. \cite{Oosterbrook1995}). The unusual
X-ray behaviour of Cir X-1 has been theorised by Oosterbrook et al
\cite{Oosterbrook1995} to arise as a result of near-supercritical
accretion onto a {\em low magnetic field} neutron star.

LSI+61$^{\circ}$ 303, like SS 433, is a weak X-ray source by
comparison with the other X-ray binaries, with a luminosity of $\sim
10^{34}$ erg s$^{-1}$ (e.g. \cite{Goldoni1995}).  X-ray flaring once
per 26.5 d orbit is interpreted as being due to supercritical
accretion during periastron passage of the compact object. High time
resolution observations reveal no evidence for X-ray pulsations in the
system \cite{Taylor1996}.

1E 1740.7-2942 \& GRS 1758-258 are relatively weak soft X-ray sources
but dominate the galactic centre region at energies above several
hundred keV \cite{Mirabel1994b}. 1E 1740.7-2942 is in fact the hardest
X-ray source within 1$^{\circ}$ of the galactic centre and may be
associated with 511 keV e$^-$e$^+$ emission, hence its popular name
`The Great Annihilator'. The 2-500 keV spectrum of 1E 1740.7-2942 is
very similar to that of the black hole candidate Cyg X-1
\cite{Churazov1994}.

\section{Discussion}

Lack of space precludes an in-depth discussion of the nature and
observational properties of the RJXRBs. While the `class' contains a
highly varied and seemingly disparate group of systems, it should be
borne in mind that it may only be the environment in the {\em inner
region of the accretion disc} which allows/forces the formation of a
radio-jet, and that jet formation may be to a large extent independent
of the properties of the individual components of the binary.

We choose to briefly highlight three properties which may be common
to (are at least {\em consistent with}) all the RJXRB :

\begin{itemize}
\item
Presence of an accretion disc during jet formation : long
required by many theoretical models, much of the spectroscopic and
photometric evidence is consistent with this scenario.
\item
Correlated radio -- X-ray behaviour : In all sources there seems
to be at least some degree of radio -- X-ray correlation, though
properties such as lags between flaring in the two regimes do vary.
\item
Low magnetic field compact object : Many of the RJXRB are convincing
black-hole candidates, illustrating that the compact object does not
supply the magnetic field required for the synchrotron emission (it
most likely originating in material in the accretion disc). The only
source for which there is direct evidence of a neutron star is Cir
X-1, a source in which the unusual X-ray properties have been modelled
as arising from a high rate of accretion onto a {\em low magnetic
field} neutron star.  Combined with a lack of detected radio emission
from {\em any} X-ray pulsar, the implication is that a strong compact
object magnetic field inhibits jet formation, possibly by preventing
formation of a stable inner accretion disc region.
\end{itemize}

\section{Conclusions}

We have presented a review of the observational properties of the
eight sources we consider to be radio-jet X-ray binaries. We note that
while the sources vary widely in systems parameters, all data is
consistent with the formation of an accretion disc during jet
formation (at least), some degree of correlated radio -- X-ray
behaviour, and a low or non-existent (i.e. black hole) compact object
magnetic field.

\ack

We would like to thank the NRL-GBI and ASM/RXTE teams for providing a
wealth of superb data to the astronomical community. We would also
like to thank all the people at the Jodrell Bank workshop, plus Dick
Hunstead, Carole Haswell, George Nicholson and many others for useful
discussions.

Basic research in radio interferometry at the Naval Research
Laboratory was supported by the Office of Naval Research.


\begin{thebibliography}{}

\bibitem{Bailyn1995}
Bailyn, C.D., Orosz, J.A., McClintock, J., Remillard, R., 1995, {\em
Nature}, {\bf 378}, 157

\bibitem{Bursov1995}
Bursov, N.N., Trushkin, S.A., 1995, {\em Astronomy Letters}, {\bf 21},
145 

\bibitem{Callanan1997}
Callanan, P., Garcia, M., 1997, in ``Accretion Phenomena and Related
Outflows.'', PASP, in press

\bibitem{CastroTirada1996}
Castro-Tirada, A.J., Geballe, T.R., Lund, N., 1996, {\em ApJ}, {\bf
461}, L99 

\bibitem{Chadwick1985}
Chadwick, P.M. {\it et al.}, 1985, {\em Nature}, {\bf 318}, 642

\bibitem{Chaty1996}
Chaty, S., Mirabel, I.F., Duc, P.A., Wink, J.E., Rodriguez, L.F.,
1996, {\em A\&A}, {\bf 310}, 825

\bibitem{Churazov1994}
Churazov, E. {\it et al.}, 1994, in Proc. XXVIIth ESLAB Symposium :
``Frontiers of Space and Ground-based Astronomy''., Kluwer, p.35

\bibitem{Ebisawa1996}
Ebisawa, K., 1996, in ``X-ray Imaging and Spectroscopy of Cosmic Hot
Plasmas'' -- International Symposium on X-ray Astronomy, in press

\bibitem{Fender1995}
Fender, R.P., Bell Burnell, S.J., Garrington, S.T., Spencer, R.E.,
Pooley, G.G., 1995, {\em MNRAS}, {\bf 274}, 633

\bibitem{Fender1996}
Fender, R.P., Bell Burnell, S.J., Williams, P.M., Webster, A.S., 1996,
{\em MNRAS}, in press

\bibitem{Fender1997a}
Fender, R.P., Bell Burnell, S.J., Waltman, E.B., Pooley, G.G., Ghigo, F.D.,
Foster, R.S., 1997a, {\em MNRAS}, in press

\bibitem{Fender1997b}
Fender, R.P., Pooley, G.G., Harmon, B.A., Robinson, C., Zhang, N.,
Canosa, C., 1997b ,in ``Accretion Phenomena and Related Outflows.'',
PASP, in press 

\bibitem{Foster1996}
Foster, R.S., Waltman, E.B., Tavani, M., Harmon, B.A., Zhang, S.N.,
Paciesas, W.S., Ghigo, F.D., 1996, {\em ApJ}, {\bf 467}, L81

\bibitem{Glass1994}
Glass, I.S., 1994, {\em MNRAS}, {\bf 268}, 742

\bibitem{Goldoni1995}
Goldoni, P., Mereghetti, S., 1995, {\em A\&A}, {\bf 299}, 751

\bibitem{Gregory1972}
Gregory, P.C. {\it et al.}, 1972, {\em Nature}, {\bf 239}, 114

\bibitem{Greiner1995}
Greiner, J., Predehl, P., Phol, M., 1995, {\em A\&A}, {\bf 297}, L67

\bibitem{Greiner1997}
Greiner, J., Remillard, R., Morgan, E., 1997, in ``Accretion Phenomena
and Related Outflows.'', PASP, in press

\bibitem{Hermsen1987}
Hermsen, W., {\it et al.}, 1987, {\em A\&A}, {\bf 175}, 141

\bibitem{Hjellming1995a}
Hjellmiing, R.M., Han, X., 1995, in Lewin, W.H.G., van Paradijs, J.,
van den Heuvel, E.P.J. (eds), X-ray Binaries., CUP, p.308

\bibitem{Hjellming1981a}
Hjellming, R.M., Johnston, K.J., 1981a, {\em ApJ}, {\bf 246}, L141

\bibitem{Hjellming1981b}
Hjellming, R.M., Johnston, K.J., 1981b, {\em Nature}, {\bf 290}, 100 

\bibitem{Hjellming1995b}
Hjellming, R.M., Rupen, M.P., 1995, {\em Nature}, {\bf 375}, 464

\bibitem{Kafuku1994}
Kafuku, S., {\it et al.}, 1994, {\em MNRAS}, {\bf 268}, 437

\bibitem{Kitamoto1994}
Kitamoto, S., Miyamoto, S., Waltman, E.B., Fiedler, R.L., Johnston, K.,
Ghigo, F.D., 1994, {\em A\&A}, {\bf 281}, L85

\bibitem{Kodaira1985}
Kodaira, K., Nakada, Y., Backman, D.E., 1985, {\em ApJ}, {\bf 296},
232 

\bibitem{Liedahl1996}
Liedahl, D.A., Paerels, F., 1996, {\em ApJ}, submitted

\bibitem{Manchanda1993}
Manchanda, R.K., 1993, {\em Adv. Space Sci. Res.}, {\bf 13}, 331

\bibitem{Margon1984}
Margon, B., 1984, {\em Annual Rev. Astron. Astrophys.}, {\bf 22}, 507 

\bibitem{Margon1980}
Margon, B., Grandi, S.A., Downes, R.A., 1980, {\em ApJ}, {\bf 241},
306 

\bibitem{Marti1993}
Mart\'{\i}, J., 1993, {\rm PhD thesis}, University of Barcelona

\bibitem{Marti1995}
Mart\'{\i}, J., Paredes, J.M., 1995, {\em A\&A}, {\bf 298}, 151

\bibitem{Mason1986}
Mason, K.O., Cordova, F.A., White, N.E., 1986, {\em ApJ}, {\bf 309},
700 

\bibitem{Massi1993}
Massi, M., Paredes, J.M., Estalella, R., Felli, M., 1993, {\em A\&A},
{\bf 269}, 249

\bibitem{Matz1996}
Matz, S.M., Fender, R.P., Bell Burnell, S.J., Grove, J.E., Strickman,
M.S., {\em A\&A}, in press

\bibitem{Mirabel1994a}
Mirabel, I.F., Rodriguez, L.F., 1994, {\em Nature}, {\bf 371}, 46

\bibitem{Mirabel1994b}
Mirabel, I.F., 1994, {\em ApJS}, {\bf 92}, 369

\bibitem{Moneti1992}
Moneti, A., 1992, {\em A\&A}, {\bf 260}, L7

\bibitem{Nicholson1980}
Nicholson, G.D., Feast, M.W., Glass, I.S., 1980, {\em MNRAS}, {\bf
191}, 293 

\bibitem{Oosterbrook1995}
Oosterbrook, T., van der Klis, M., Kuulkers, E., van Paradijs J. and
Lewin, W.H.G., 1995, {\em A\&A}, {\bf 297}, 1410

\bibitem{Paredes1994}
Paredes, J.M., {\it et al.}, 1994, {\em A\&A}, {\bf 288}, 519

\bibitem{Pooley1996a}
Pooley, G.G., 1996a, {\it I.A.U. Circ 6269}

\bibitem{Pooley1996b}
Pooley, G.G., 1996b, {\it I.A.U. Circ 6411}

\bibitem{Predehl1995}
Predehl, P., Schmitt, J.H.M.M., 1995, {\em A\&A}, {\bf 293}, 889

\bibitem{Priedhorsky1986}
Priedhorsky, W., Terrel, J., 1986, {\em ApJ}, {\bf 301}, 886

\bibitem{Ray1996}
Ray, P.S., Foster, R.S., Waltman, E.B., Ghigo, F.D., Tavani, M., 1996,
{\em ApJ}, submitted

\bibitem{Sams1996}
Sams, B.J., Eckart, A., Sunyaev, R., 1996, {\it Nature}, {\bf 382}, 47

\bibitem{Schalinski1995}
Schalinski, C.J., Johnston, K.J., Witzel, A., {\it et al.}, 1995, {\em
ApJ}, {\bf 447}, 752

\bibitem{Spencer1979}
Spencer, R.E., 1979, {\em Nature}, {\bf 282}, 483

\bibitem{Stewart1993}
Stewart, R.T., Caswell, J.L., Haynes, R.F., Nelson, G.J., 1993, {\em
MNRAS}, {\bf 261}, 593

\bibitem{Taylor1991}
Taylor, A.R., Gregory, P.C., Duric, N., Tsutsumi, T., 1991, {\em
Nature}, {\bf 351}, 547

\bibitem{Taylor1996}
Taylor, A.R., Young, G., Peracaula, M., Kenny, H.T., Gregory, P.C.,
1996, {\em A\&A}, {\bf 305}, 817

\bibitem{Thompson1979}
Thompson, R.I., Rieke, G.H., Tokunaga, A.T., Lebofsky, M.J., 1979,
{\em ApJ}, {\bf 234}, L135

\bibitem{Tingay1995}
Tingay, S.J., Jauncey, D.L., Preston, R.A., Reynolds, J.E., Meier,
D.L., {\it et al.}, 1995, {\em Nature}, {\bf 374}, 141

\bibitem{Tsutsumi1996}
Tsutsumi, T., Peracaula, M., Taylor, A.R., 1996, in ``Radio emission
from the Stars and the Sun'', Taylor, A.R. and Paredes, J.M. (eds),
ASP Conference Series, Vol. 93, 258

\bibitem{vKerkwijk1992}
van Kerkwijk, M.H. {\it et al.}, 1992, {\em Nature}, {\bf 355}, 703 

\bibitem{Waltman1994}
Waltman, E.B., Fiedler, R.L., Johnston, K.J., Ghigo, F.D., 1994, {\em
AJ}, {\bf 108}, 179

\bibitem{Wang1990}
Wang, Z.R., McCray, R., Chen, Y., Qu, Q.Y., 1990, {\em A\&A}, {\bf
240}, 98 

\bibitem{Watson1983}
Watson, M.G., Willingale, R., Grindlay, J.E., Seward, F.D., 1983, {\em
ApJ}, {\bf 273}, 688

\end{thebibliography}
\end{document}